\documentclass[twocolumn]{aastex631}

\usepackage{amsmath}
\usepackage{hyperref}
\usepackage{array}
\usepackage{threeparttable}

\usepackage{savesym}
\savesymbol{tablenum}
\usepackage{siunitx}
\restoresymbol{SIX}{tablenum}

\newcommand{\Msun}{M_{\odot}}
\newcommand{\Lsun}{L_{\odot}}

\newcommand{\HII}{H{\sc ii}}

\newcommand{\kms}{\mbox{${\rm km~s^{-1}}$}}
\newcommand{\kmsMpc}{\mbox{km s$^{-1}$ Mpc$^{-1}$}}
\newcommand{\MC}{\multicolumn}

\usepackage{graphicx} 

\begin{document}

\title{KKR\,18: A Late Arrival Galaxy?}

\author[0000-0002-9291-1981]{R. Brent Tully}
\email{tully@ifa.hawaii.edu}
\affil{Institute for Astronomy, University of Hawaii, 2680 Woodlawn Drive, Honolulu, HI 96822, USA}

\author[0009-0007-3326-4191]{Valentina E. Karachentseva}
\affiliation{Main Astronomical Observatory, National Academy of Sciences of Ukraine, Kiev, 03143, Ukraine}

\author[0000-0003-0307-4344]{Igor D. Karachentsev}
\affiliation{Special Astrophysical Observatory, Russian Academy of Sciences, Nizhniy Arkhyz, Karachai-Cherkessia 369167, Russia}

\author[0000-0002-5259-2314]{Gagandeep S.\ Anand}
\affiliation{Space Telescope Science Institute, 3700 San Martin Drive, Baltimore, MD 21218, USA}

\author[0009-0004-4126-8924]{Maksim I.\ Chazov}
\affiliation{Special Astrophysical Observatory, Russian Academy of Sciences, Nizhniy Arkhyz, Karachai-Cherkessia 369167, Russia}

\author[0000-0001-6927-5473]{Serafim S. Kaisin}
\affiliation{Special Astrophysical Observatory, Russian Academy of Sciences, Nizhniy Arkhyz, Karachai-Cherkessia 369167, Russia}

\author[0000-0002-5514-3354]{Ehsan Kourkchi}
\email{ehsan@ifa.hawaii.edu}
\affil{Institute for Astronomy, University of Hawaii, 2680 Woodlawn Drive, Honolulu, HI 96822, USA}

\author[0000-0003-0736-7609]{Lidia N. Makarova}
\affiliation{Special Astrophysical Observatory, Russian Academy of Sciences, Nizhniy Arkhyz, Karachai-Cherkessia 369167, Russia}

\author[0000-0002-4707-4467]{Simon A. Pustilnik}
\affiliation{Special Astrophysical Observatory, Russian Academy of Sciences, Nizhniy Arkhyz, Karachai-Cherkessia 369167, Russia}

\author[0000-0002-3234-8699]{Edward J. Shaya}
\affiliation{Astronomy Department, University of Maryland, College Park, MD 20743, USA}

\author[0000-0003-1303-6996]{Arina L. Tepliakova}
\affiliation{Special Astrophysical Observatory, Russian Academy of Sciences, Nizhniy Arkhyz, Karachai-Cherkessia 369167, Russia}

\begin{abstract}
KKR\,18 is a dwarf irregular galaxy in the Local Void at a distance of 9.46~Mpc.  It is gas rich and was experiencing vigorous star formation $20-100$~Myr ago.  Its metallicity is 4\% of Solar from an emission spectrum, qualifying it as extremely metal poor (XMP).  Its most remarkable characteristic is an extremely deficient Red Giant Branch.  This feature is shared with a small number of other dwarf galaxies and it is inferred that they lack ancient stellar components.  It is suggested that these are late arrival galaxies (LAG).
\end{abstract}

\section{Introduction}
\label{sec:intro}

The galaxy KKR\,18 \citep{1999A&AS..135..221K} was observed in the course of Hubble Space Telescope (HST) SNAP program GO-18070 \citep{2025hst..prop18070T} because a heliocentric velocity of $720\pm3$~\kms\ \citep{2018ApJ...861...49H} indicates it is relatively close-by.
The galaxy resolves into an abundance of young stars while the Red Giant Branch (RGB) is detected but poorly populated.
This situation is an anomaly; normally the RGB is a dominant feature in a color-magnitude diagram (CMD).
For example, see the montage of CMD for Local Group galaxies in \citet{2014ApJ...789..147W} or those in the collection of CMD by \citet{2021AJ....162...80A}.
In all well-studied cases, galaxies contain old stars with ages $\sim 13$~Gyr \citep{2014ApJ...789..147W, 2015ApJ...812..158M, 2017ApJ...837..102S} consistent with the standard idea that star formation begins in halos before the epoch of reionization.
In very low mass halos star formation could be halted by the dispersal of heated gas upon reionization \citep{1996ApJ...465..608T, 2002ApJ...572L..23S}.  In any event, whether star formation is halted or continuing, the RGB is a robust feature.

The insubstantial RGB in the case of KKR\,18 raises the possibility that star formation in this galaxy began in substance only in the last few Gyr.
It is a notable fact that the galaxy lies in isolation in the Local Void.
KKR\,18 is a $B=16.1$~mag dwarf irregular galaxy with alternate names CGCG\,106-29, MCG\,+03-39-019, AGC\,252478 and PGC\,54607. 
It has the coordinates equatorial J2000 15h17m54.56s+16d18m40.0s, Galactic 22.89695 +54.45371, supergalactic 107.79259 +37.98683.

\begin{figure}[!]
\centering
\includegraphics[width=1.00\linewidth]{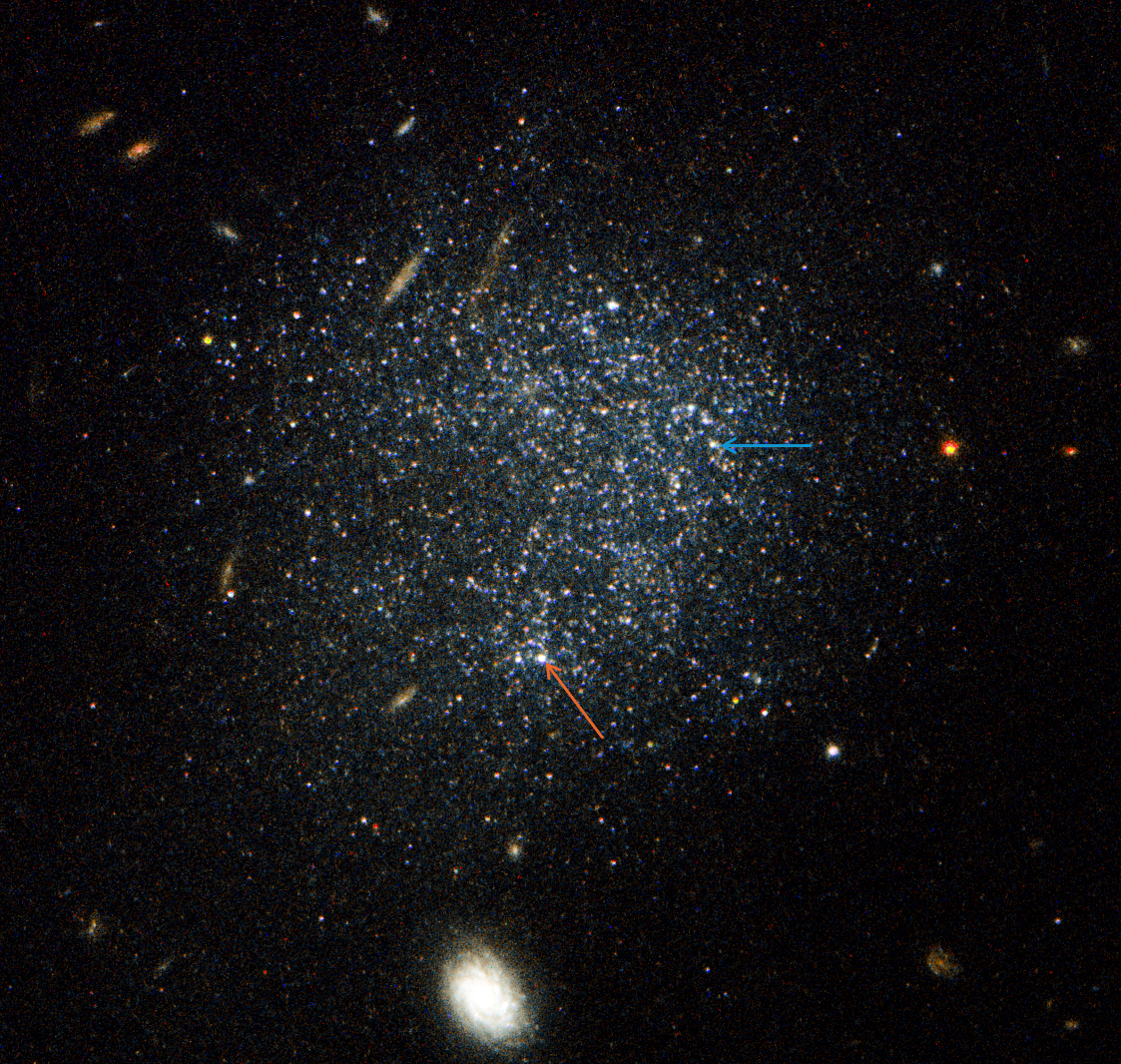}
\caption{Composite image of KKR\,18 generated from HST exposures in $F606W$ and $F814W$ ($\sim$0.85$'$$\times$0.85$'$). Blue and red arrows identify the locations of an HII region and a star cluster each discussed in the text. North is up and east is to the left.}
\label{fig:image}
\end{figure}

\section{HST Observation}
\label{sec:hstobs}

\begin{figure}[!]
\centering
\includegraphics[width=1.00\linewidth]{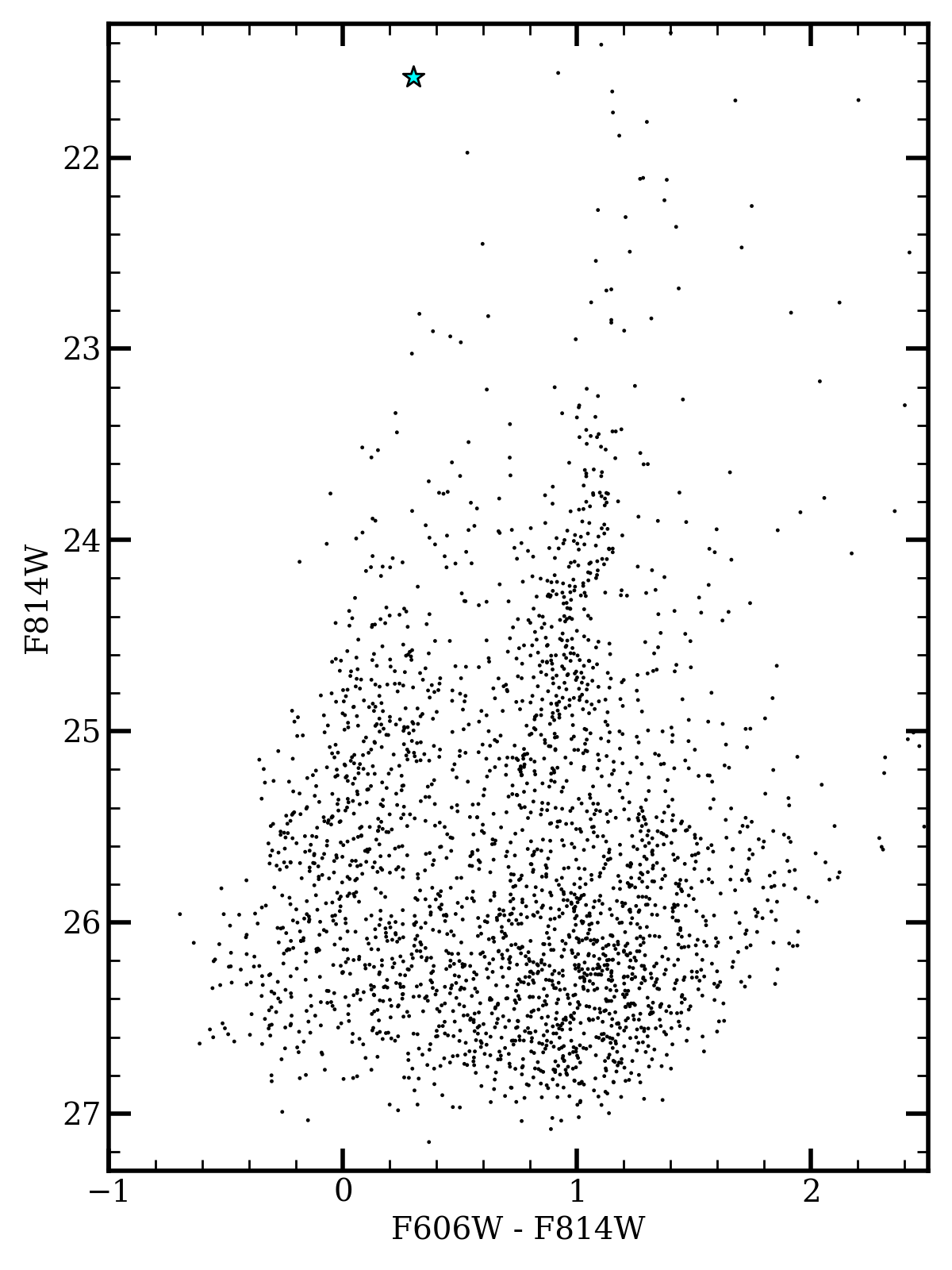}
\caption{$F814W$ vs $F606W-F814W$ CMD of the full field of KKR\,18. The star cluster discussed in the text is located by the star.}
\label{fig:fullcmd}
\end{figure}

KKR\,18 was observed with HST on 28 April, 2026, for 748 seconds with both the $F606W$ and $F814W$ filters using the Advanced Camera for Surveys (ACS). A color image of the galaxy is seen in Fig.~\ref{fig:image}. Stellar point-spread function (PSF) photometry was carried out using the DOLPHOT software package \citep{2016ascl.soft08013D}.
The PSF photometry was performed on the individual \texttt{FLC} exposures using the combined F814W \texttt{DRC} image as a reference frame for alignment of the individual underlying exposures. The reduction was carried out using DOLPHOT setup parameters recommended for the ACS camera by the PHAT/PHATTER team \citep{2021ApJS..253...53W}. The initial source list was trimmed using quality cuts to retain only sources of high quality \citep{2017AJ....154...51M, 2021AJ....162...80A}. Specifically, we retained point sources which passed the following quality cuts: 1) \texttt{Crowd}\textsubscript{F606W} + \texttt{Crowd}\textsubscript{F814W} $<$ 0.8, 2) (\texttt{Sharp}\textsubscript{F606W} + \texttt{Sharp}\textsubscript{F814W})$^{2}$ $<$ 0.075, 3) \texttt{Type} $\le$ 1, 4) $S/N_{F814W}$ $\ge$ 4, 5) $S/N_{F606W}$ $\ge$ 3, and 6) \texttt{Flag} $\le$ 2. The color-magnitude diagram (CMD) for the full field is shown in Fig.~\ref{fig:fullcmd}. Artificial star experiments are also performed within DOLPHOT to determine the levels of photometric bias, completeness, and errors. These results provide necessary input for distance and star-formation measurements.

\begin{figure}[]
\centering
\includegraphics[width=1.00\linewidth]{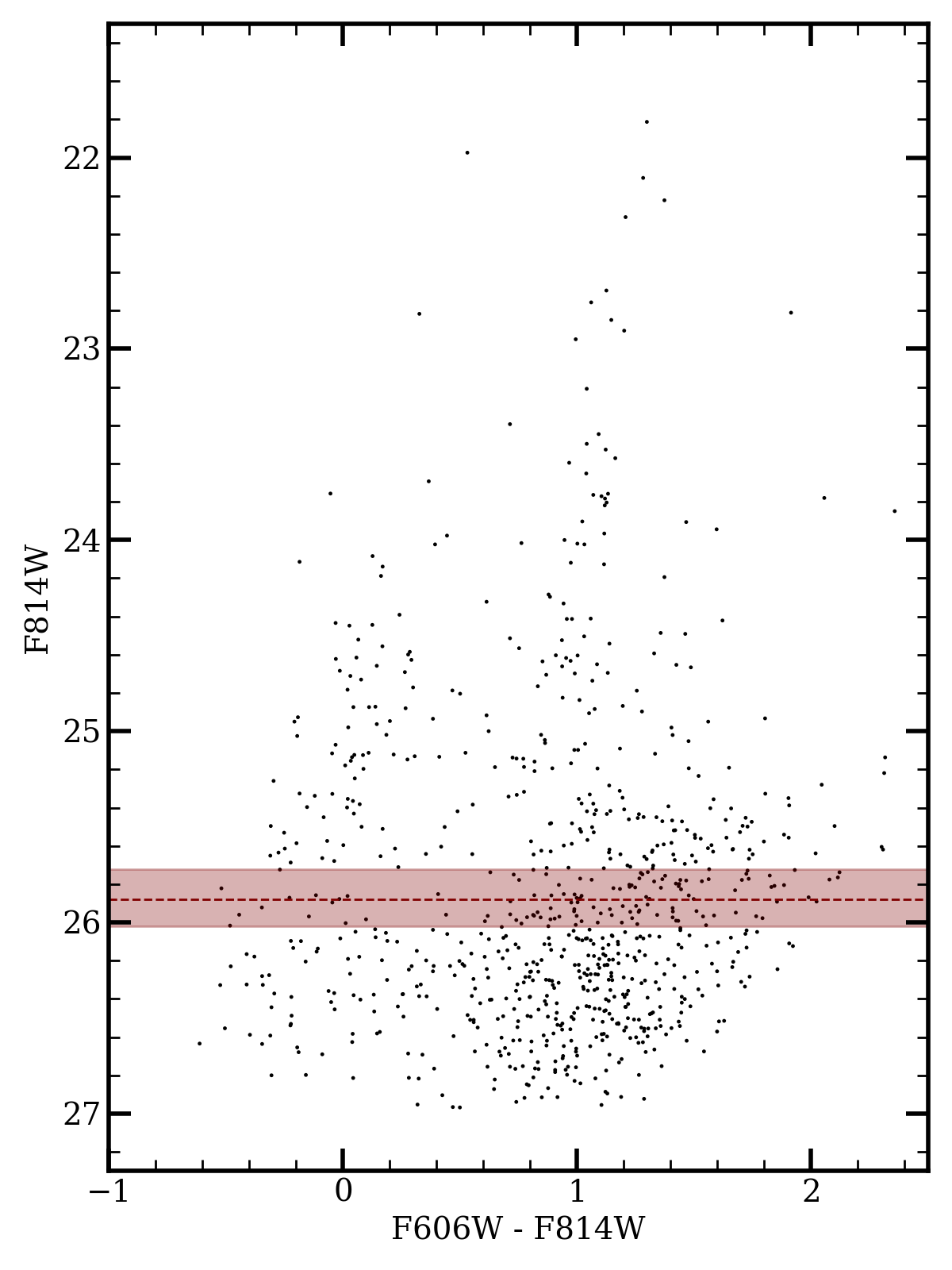}
\caption{CMD excluding central region dense with young stars. The TRGB level and uncertainty are indicated by the red dashed line and shaded region.}
\label{fig:trgbcmd}
\end{figure}

\section{Distance}
\label{sec:distance}

The trained eye can identify the probable RGB in Fig.~\ref{fig:fullcmd} but it is ambiguous in the CMD of the entire galaxy.
The more prominent features are the Red Supergiant Branch rising brightward of $F814W=25$~mag at $F606W-F814W\sim1$~mag, the Main Sequence and Blue Loop stars to the blue of $F606W-F814W\sim0.5$~mag, and Asymptotic Giant Branch stars brighter than $F814W\sim26$~mag and to the red of $F606W-F814W\sim1$~mag.

With the CMD of Fig.~\ref{fig:trgbcmd}, stars from the inner core of the galaxy are excluded. The RGB is sufficiently revealed to allow a measurement of the Tip of the RGB (TRGB), a method grounded in the standard candle nature of the brightest red giant branch stars \citep{1993ApJ...417..553L, 1996ApJ...461..713S}. 
The TRGB measurement methodology of \cite{2014AJ....148....7W} was applied in a color band $0.5<F606W-F814W<1.6$~mag including the input from recovery of the artificial stars.  The TRGB is located 
at $F814W=25.88$~mag, albeit with considerable uncertainty of $\pm0.16$~mag.
The observed magnitude is diminished by foreground reddening $A_{F814W}=0.05$~mag \citep{2011ApJ...737..103S}, with obscuration internal to the dwarf galaxy assumed to be negligible. 
The absolute value of the TRGB is taken from the calibration by \citet{2007ApJ...661..815R} adjusted to the Cosmicflows 4 scale \citep{2023ApJ...944...94T} of $M_{F814W}=-4.05$~mag which includes a small color term to account for variations induced by varying metallicities.
The derived distance modulus $m-M=29.88\pm0.18$~mag corresponds to the distance $d=9.46\pm0.79$~Mpc. 

\begin{figure}[]
\centering
\includegraphics[width=1.00\linewidth]{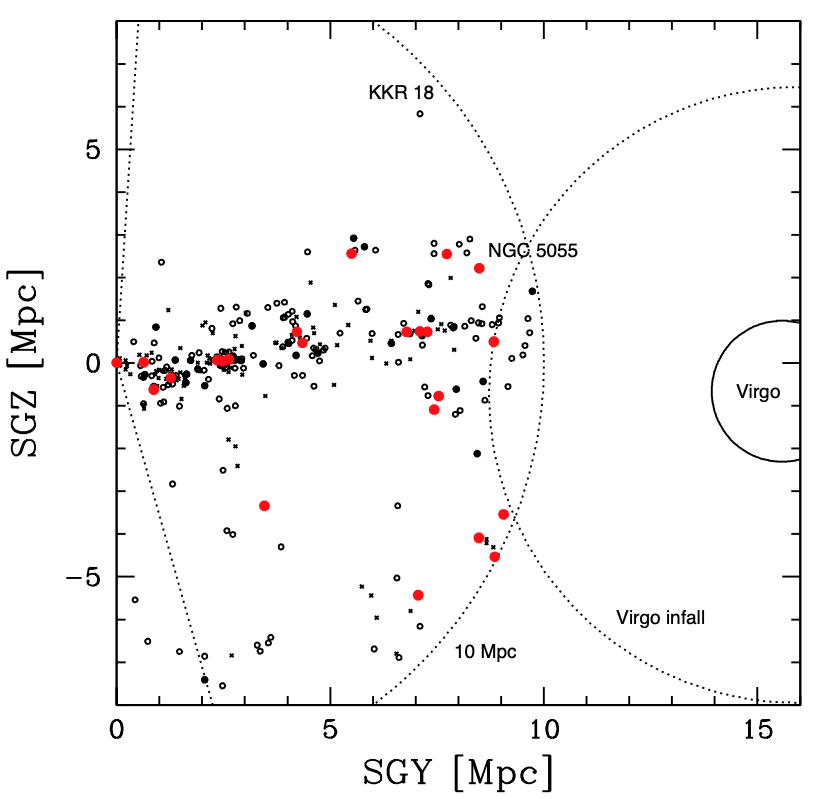}
\caption{Location of galaxies from the 10 Mpc volume sample in supergalactic SGY-SGZ coordinates. The depth of the display is $-10<\rm{SGX}<+4$ Mpc. Prominent galaxies are shown in red and successively smaller galaxies in black filled and open symbols and crosses.  The virial region of the Virgo Cluster lies within the solid circle while the dotted circle about it shows the domain of infall to Virgo.  The Milky Way Galaxy lies at the origin with radiating dotted lines at Galactic latitude $+10^{\circ}$ and a dotted semi-circle defines the limit of the 10~Mpc volume.  The dwarf galaxy KKR\,18 is identified within the realm of the Local Void.}
\label{fig:yzLV}
\end{figure}

\section{Location in the Local Void}
\label{sec:LV}

With the measured distance, KKR\,18 has the supergalactic coordinates SGX=$-2.27$~Mpc, SGY=7.10~Mpc, SGZ=5.83~Mpc.
It's position is shown in the SGY-SGZ plane in Fig.~\ref{fig:yzLV} along with galaxies from the 10~Mpc sample of Tully et al. (submitted 2026).  The relationship of the sample to the Virgo Cluster is shown.  There is an evident concentration of galaxies to the Local Sheet \citep{2008ApJ...676..184T}.  KKR\,18, at SGZ=6~Mpc, lies above the Local Sheet within the Local Void. Only two other galaxies have TRGB distances that locate them in the Local Void, both south of the Galactic plane at negative SGY: ESO\,461-36 \citep{2011AJ....141..204K} and ALFAZOA~J1952+1428 \citep{2011ApJ...739L..26M}.
The nearest large galaxy to KKR\,18 is NGC\,5055 at a distance of 6~Mpc.

\begin{figure}[]
\centering
\includegraphics[width=1.00\linewidth]{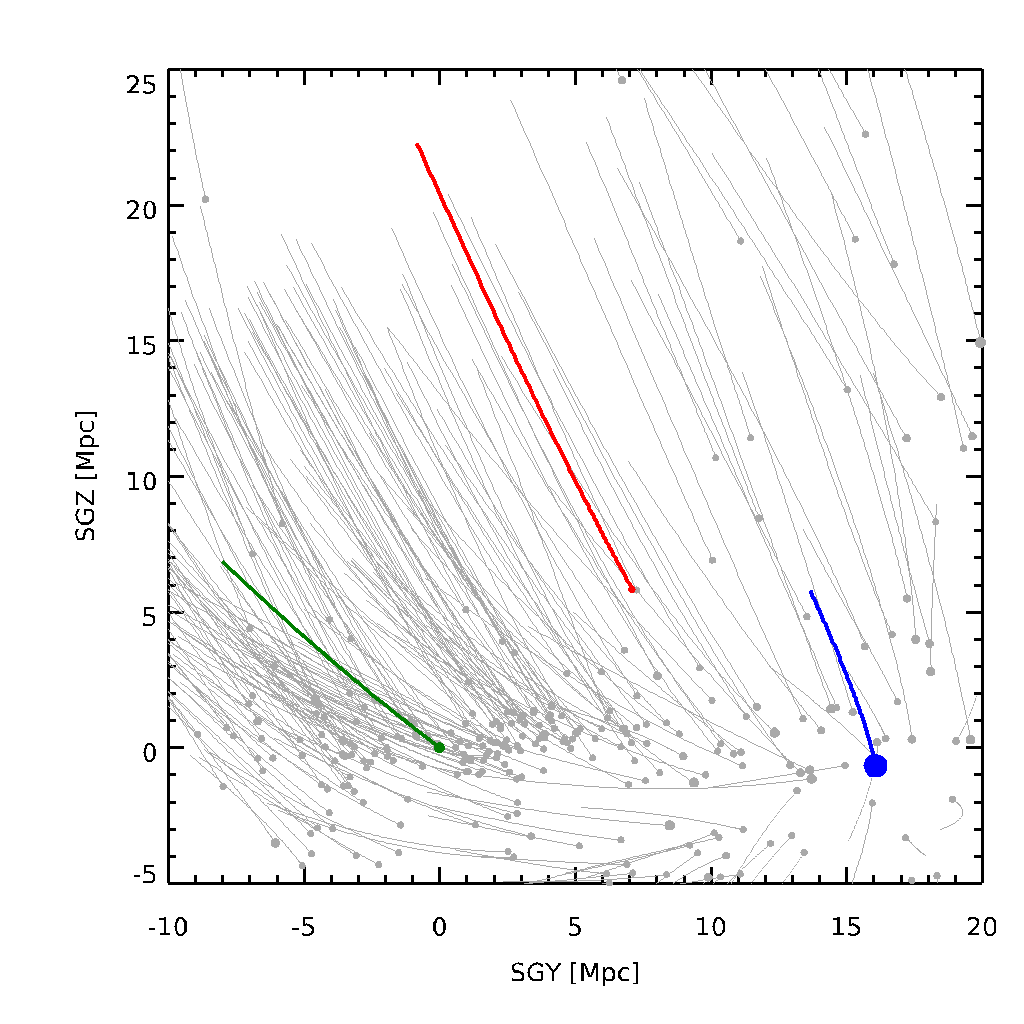}
\caption{NAM orbits of galaxies and groups from z=4 to z=0. The KKR\,18 orbit is in red, that of Milky Way is in green and Virgo Cluster is in blue.}
\label{fig:yzorb}
\end{figure}

The heliocentric velocity for KKR\,18 of 720~\kms\ translates to 771~\kms\ in the Local Sheet frame \citep{2008ApJ...676..184T} and 876~\kms\ in the CMB frame.  Assuming $H_0=75$~\kmsMpc, the galaxy has a peculiar velocity in the local reference frame of $+61$~\kms. 
Fig.~\ref{fig:yzorb} illustrates orbits of galaxies and groups determined by Numerical Action Methods (NAM) in the study by \citet{2022ApJ...927..168S}. The orbit followed by the center of mass of material that consolidated into KKR\,18 is shown in red.

Essentially all galaxies at positive SGZ are heading toward the supergalactic south, away from the Local Void. KKR\,18 is still in the void and has a large downward peculiar velocity, curling toward the Virgo Cluster. 

\section{Properties of the Resolve Stellar Population of KKR\,18}
\label{sec:isochrones}
The selection criteria for high-quality stars and the signal-to-noise thresholds were conservative. Nevertheless, the relatively short exposures of 748s per filter suffices to yield a color–magnitude diagram resolving stars spanning approximately 5.5 magnitudes. A young, bright stellar population clearly dominates the resolved stars in the CMD. We identify blue supergiants -- the upper main sequence — and stars of the blue loop with colors of roughly $\leq$ 0.6 mag. The red supergiant branch is well-defined brighter than about F814W$\simeq$25 mag and redder than $F606W-F814W\simeq0.6$ mag.  The CMD therefore unambiguously indicates that the object is a dwarf irregular galaxy. 

\begin{figure}[]
\centering
\includegraphics[width=1.00\linewidth]{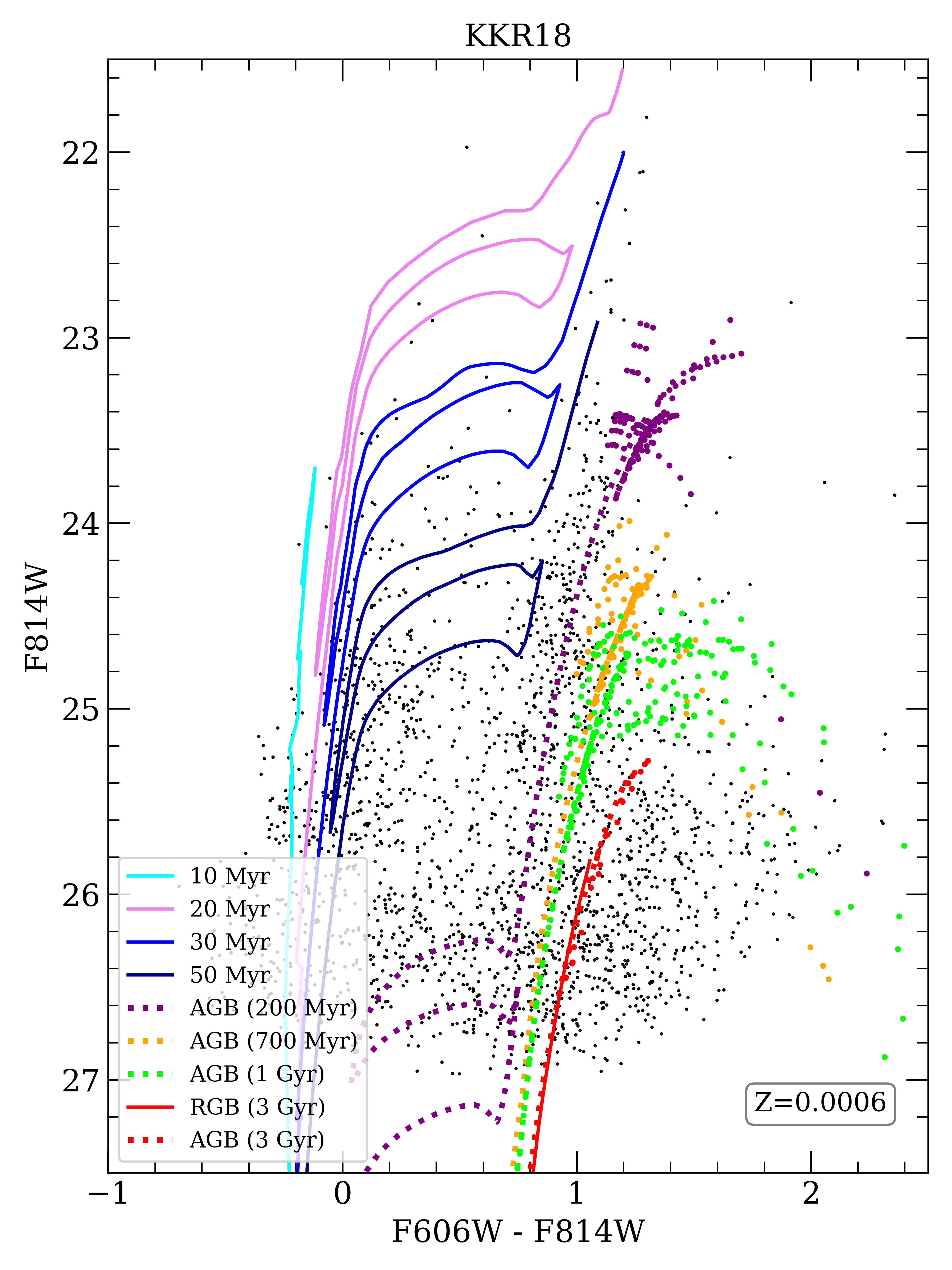}
\caption{CMD of KKR\,18 with selected PARSEC stellar isochrones superimposed 
\citep{2012MNRAS.427..127B, 2013MNRAS.434..488M}
assuming $d=9.5$~Mpc and metallicity Z=0.0006.}
\label{fig:isocmd}
\end{figure}



\subsection{Populations younger than 1 Gyr}
\label{youngstars}

The measured gas phase abundance of 0.04 Solar (see Sect.~\ref{sec:oxygen}) provides an approximate upper limit for the stellar population metallicity in the galaxy, assuming in situ enrichment of the gas. We superimposed
Padova isochrones of the respective metallicity of Z=0.0006 over the CMD of KKR\,18 to estimate possible ages of the resolved stars. As seen in Fig.~\ref{fig:isocmd}, 
blue supergiants on the main sequence rise to $F814W\sim24$ mag corresponding to stars of age about 10 Myr and maximum masses of about $2.6~\Msun$.

From the fitted isochrones there is an evident rich population of blue and red supergiant stars with ages of about 20--100 Myr, as well as red stars at the early AGB stage at about 200--700 Myr. 
The brightest supergiants have $F814W$ magnitudes roughly between 22 and 23 mag. If these stars 
are indeed members of KKR\,18, their ages are of order 15--30 Myr and 
their masses are about 10--13~$M_\odot$.

Thus, the analysis of the CMD indicates that, when the stellar metallicity is 
constrained by the gas-phase metallicity measured in the star-forming regions, 
the present-day star formation activity (within the last $\sim$10 Myr) is 
relatively weak. The most luminous stars appear to be older and more massive. 
Taken together, these findings are consistent with a scenario in which the 
current star formation activity is gradually fading.

\subsection{Populations older than 1 Gyr}
\label{oldstars}



It is evident that RGB stars do exist.  However there is little constraint on ages from the RGB that could run from 2 to 13 Gyr. Observational uncertainties are substantial in photometry at the level of the RGB that is within the last magnitude above the adopted $S/N$ limits at $F814W\sim26.8$ mag and $F606W\sim27.8$ mag. 
By contrast, AGB stars redward of $F606W-F814W \sim 1$ mag and with $F814 <26$ mag are detected at reliable photometry levels and are abundant.

\begin{figure}[]
\centering
\includegraphics[width=1.00\linewidth]{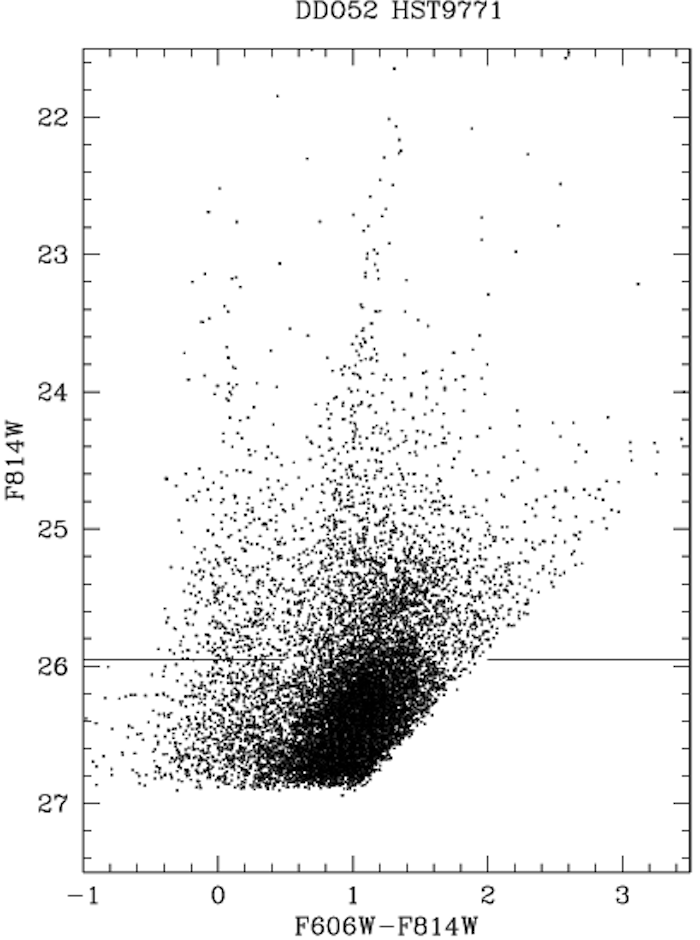}
\caption{CMD for DDO\,52, distance 9.62~Mpc, extracted from the CMDs/TRGB catalog in EDD.}
\label{fig:ddo52}
\end{figure}

\begin{figure}[]
\centering
\includegraphics[width=1.00\linewidth]{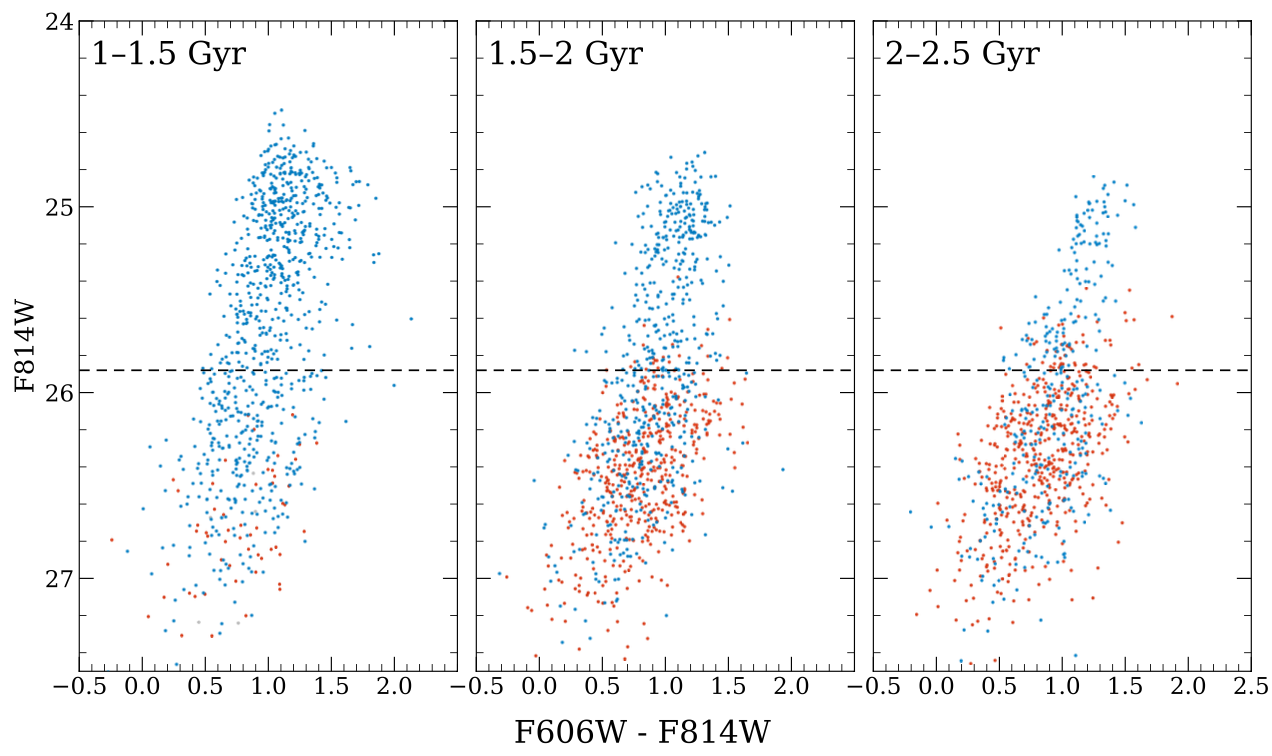}
\caption{CMD of synthetic stellar populations with $Z=0.0004$, fixed masses, and 3 age intervals encompassing 1 Gyr to 2.5 Gyr.  We used PARSEC stellar library, photometric errors and completeness distribution of KKR\,18 and Salpeter IMF. RGB stars are in red and AGB stars are blue.  The dashed horizontal line is at the measured TRGB magnitude of $F814W=25.88$ mag.}
\label{fig:rgb/agb}
\end{figure}

We must be clear about the dilemma posed by the KKR\,18 CMD.  The CMD of a galaxy of comparable luminosity and similar young star activity at about the same distance is shown in Fig.~\ref{fig:ddo52}.  There are familiar features to those seen in the Fig.~\ref{fig:fullcmd} CMD of KKR\,18 but the prominent RGB in the case of DDO\,52 is a dramatic difference.  One can troll through hundreds of cases in the CMDs/TRGB catalog in the Extragalactic Distance Database \citep{2009AJ....138..323T, 2021AJ....162...80A} to be convinced of the ubiquitousness of the dominance of the RGB feature.

We are to explain two observations: the unusually diminished RGB and the substantial AGB.  Fig.~\ref{fig:rgb/agb} informs this discussion.  Synthetic CMD have been constructed in the course of modeling the star formation history represented through isochrones in Fig.~\ref{fig:isocmd}.
In each case, the same metallicity (Z=0.0004) and galaxy mass ($3\times10^7~\Msun$) are taken.  The three panels show the synthetic CMD for populations in three adjacent ages.  In the 1 to 1.5 Gyr interval, the RGB is just starting to build while there is a rich AGB (EAGB+TPAGB) rising above the level of the fitted TRGB.  At ages older than 1.5~Gyr RGB stars are beginning to reach the level of the TRGB fit while the AGB population relatively diminishes with age.  
The ratio AGB/RGB is shown as functions of age and metallicity in Fig.~15 of \citet{2011AJ....141..106J}.
With increasing ages, less massive stars reach the RGB, drawing from the initial mass function to increase exponentially in numbers.

The implications are, first, that there was substantial star formation about 1~Gyr ago giving the observed AGB population and, second, that there was star formation 2-3~Gyr ago generating the small RGB population.  Star formation at much earlier times would have left a signature in the numbers of RGB stars that is not seen.

\section{BTA observations}
\label{sec:BTA}

KKR\,18 was observed at the BTA-6m telescope at the Special Astrophysical Observatory in the frameworks of two programs. The first program aims (in synergy with SALT, the South African Large Telescope) to conduct an unbiased spectral survey of a subsample of about 240 void galaxies from the Nearby Void Galaxy sample \citep{2019MNRAS.482.4329P}, which fall within the Local Volume (LV, that is reside at distances of $\lesssim$11~Mpc from the Local Group center).

The main goal of the BTA program is to collect data on galaxy gas metallicities to address their evolutionary status. Part of the SALT results are published in \citet{2026MNRAS.548ag559P}, while the BTA part will appear in a forthcoming paper by Pustilnik et al. KKR\,18, as a representative of this subsample, was observed to obtain its oxygen abundance.
The second program is related to a large project aiming to study star formation of the LV galaxies via imaging of their H$\alpha$ emission discussed next.  
As one of the LV late-type dwarfs, KKR\,18 was observed to map its H$\alpha$ and to obtain its total H$\alpha$ luminosity.

\subsection{H$\alpha$ imaging}
\label{ssec:Ha}

\begin{figure}[]
\centering
\includegraphics[width=1.00\linewidth]{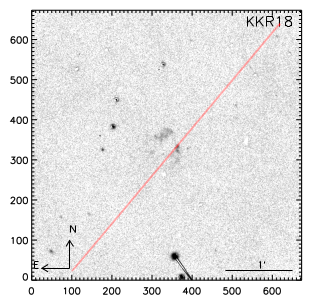}
\caption{Image of the KKR\,18 field in an H${\alpha}$ filter centered at $\lambda6560~\mathring{A}$ and width $\Delta\lambda=73~\mathring{A}$. The orientation of the slit of the spectroscopic observation is shown by the red line.}
\label{fig:Halpha}
\end{figure}

H$\alpha$ imaging was obtained in the course of an H$\alpha$ imaging survey of nearby dwarf galaxies \citep{2012MNRAS.425.2083K} with the next generation multi-mode focal reducer SCORPIO-2 \citep{2011BaltA..20..363A} 
at the prime focus of BTA on June 11, 2026.  Two exposures with seeing 1.3$''$ totaling 20~min were acquired in a $\Delta\lambda=73~\mathring{A}$ FWHM H${\alpha}$ filter and differenced with exposures in the continuum bracketing H${\alpha}$.


Fig.~\ref{fig:Halpha} is an image of the field of KKR\,18 that isolates the H${\alpha}$ flux with the imposed position of the slit of the spectroscopic observation.  For all the evidence of young stars in the CMD, the evidence in nebular emission for current star formation (stars with masses $>10~\Msun$ and ages $<20$~Myr) is sparse.
The H$\alpha$ flux across the extent of the galaxy is 
F(H$\alpha)=8.36\pm0.63 ~10^{-15}$ erg cm$^2$ s$^{-1}$.
Following \citet{1998ARA&A..36..189K}, the star formation rate (SFR) is
log(SFR$_{H\alpha}$) = 8.98 + log(F$_{H\alpha})$c + 2log$d = -3.13$  yr$^{-1}$ with $d=9.46$~Mpc and including the correction log(F$_{H\alpha})=+0.02$ accounting for Galactic extinction.
This SFR is an order of magnitude less than that inferred from the far ultraviolet (FUV) flux.
Following \citet{2011ApJS..192....6L}, flux from the GALEX FUV filter translates to SFR$_{FUV}$ as
log(SFR$_{FUV}) = 2.78-0.4 m_{FUV} + 2{\rm log} d$.
With the apparent FUV magnitude $m_{FUV}=17.20$ and a correction for Galactic extinction of 0.20 mag, then
log(SFR$_{FUV})= -2.07$~yr$^{-1}$. 
Evidently, star formation was more vigorous $20-50$~Myr ago, consistent with the CMD analysis of Section~{\ref{sec:isochrones}.
This situation with SFR$_{H\alpha}$ much less than SFR$_{FUV}$ is not particularly uncommon.  \citet{2021MNRAS.506.1346K} identified two dozen nearby late-type galaxies with SFR$_{H\alpha}$/SFR$_{FUV}<1/20$. SFR$_{H\alpha}$ describes the immediate star formation activity whereas SFR$_{FUV}$ describes the star formation condition over extended recent history.

\subsection{Spectral observations}
\label{ssec:spectra}

Spectral observations of KKR\,18 were conducted with the multi-mode focal reducer SCORPIO \citep{2005AstL...31..194A} at the prime focus of the SAO 6-m telescope (BTA) on February 21, 2026 in the long-slit mode. Slit width and length are 1.2 arcsec and 6~arcmin, respectively.
Two VPHG gratings 1200B and 1200R were used with full ranges of 3650--5450~\AA\ and 
5680--7430~\AA, respectively, with FWHM = 5.5~\AA. The slit for both gratings was in exactly the same position.  
The CCD 2K$\times$2K detector was EEV-42-40. The pixel size was 13.5$\mu$, corresponding to 0.18\arcsec\ on the detector. The binning factor of 2 was used along the slit. To minimize the effect of atmospheric dispersion on the relative line fluxes \citep{1982PASP...94..715F}, the slit was oriented close to the current parallactic angle. For the absolute flux calibration, the spectrophotometric standard star G191-B2B, was observed during the same night.

The transparency during the night was variable. Therefore, effectively only three exposures of 600s each were used in the blue and one in the red.  
The slit was pointed at the brightest H$\alpha$ region as seen on the BTA H$\alpha$ image in Fig.~\ref{fig:Halpha}.
Its location in the HST color image of Fig.~\ref{fig:image} is indicated by the blue arrow.  It is the faint object slightly W of a brighter star.

The primary reduction was done with the use of {\it IRAF}\footnote{IRAF: the Image Reduction and Analysis Facility is distributed by the National Optical Astronomy Observatory, which is operated by the Association of Universities for Research in Astronomy, Inc. (AURA) under cooperative agreement with the National Science Foundation (NSF).}
and {\it MIDAS}\footnote{MIDAS is an acronym for the European Southern
Observatory package -- Munich Image Data Analysis System.}. Procedures are described in detail in \citet{2016MNRAS.463..670P}. 
The standard pipeline included removal of cosmic ray hits, bias subtraction, flat-field correction,
wavelength calibration and night-sky background subtraction. With the use of the spectrophotometric standard data, all
spectra were transformed to absolute fluxes. The individual 2D spectra were aligned and combined. Separate
the blue and the red spectra of the only \HII\ region were then extracted by summing-up without
weighting of 12 rows along the slit ($\sim$4\arcsec). 
The resulting 1D spectrum, combining the blue and red parts, is shown in Fig.~\ref{fig:spectrum} (top panel).


\begin{figure}[]
\centering
\includegraphics[width=7.0cm,angle=-90,clip=]{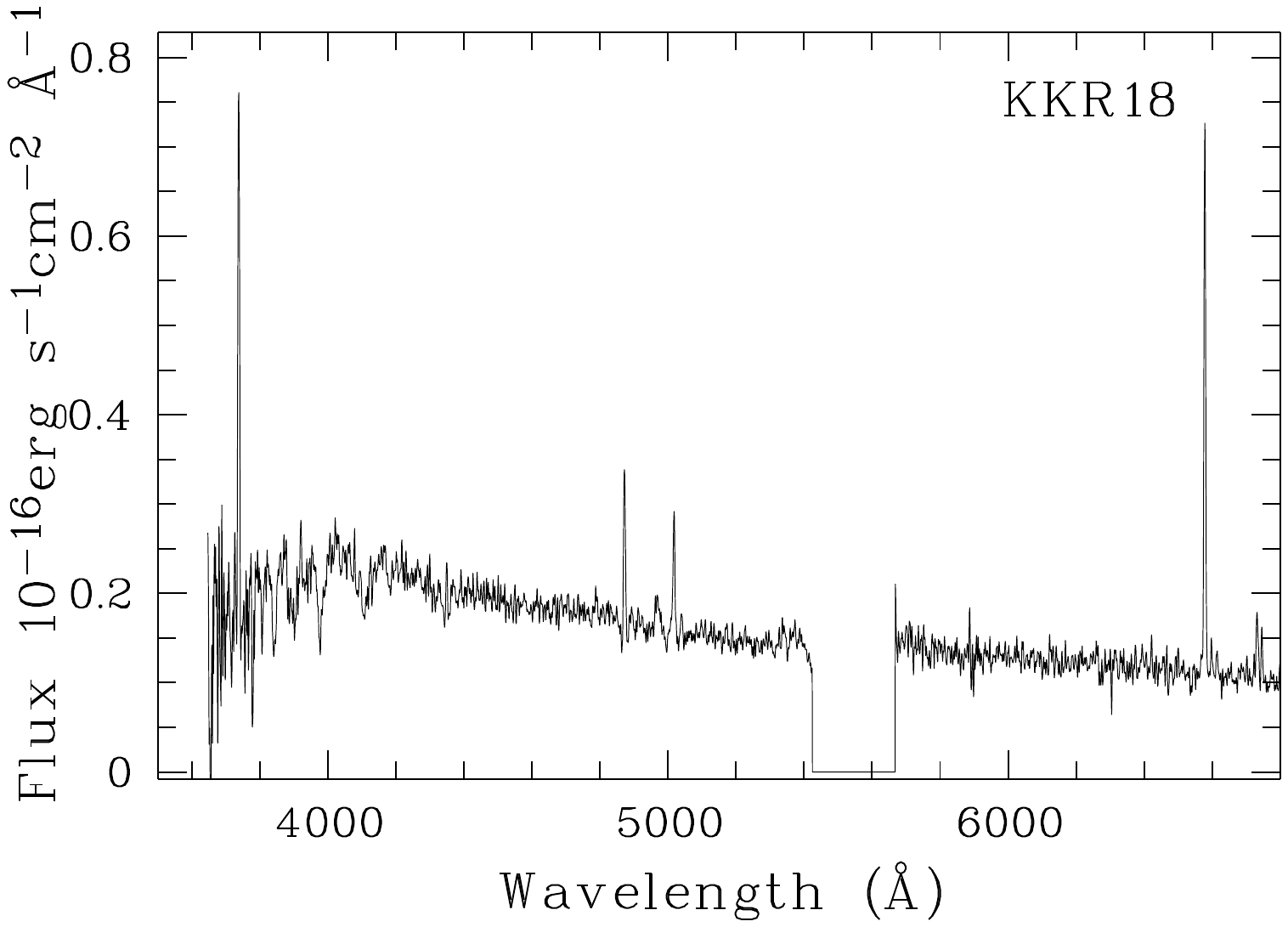}
\includegraphics[width=7.0cm,angle=-90,clip=]{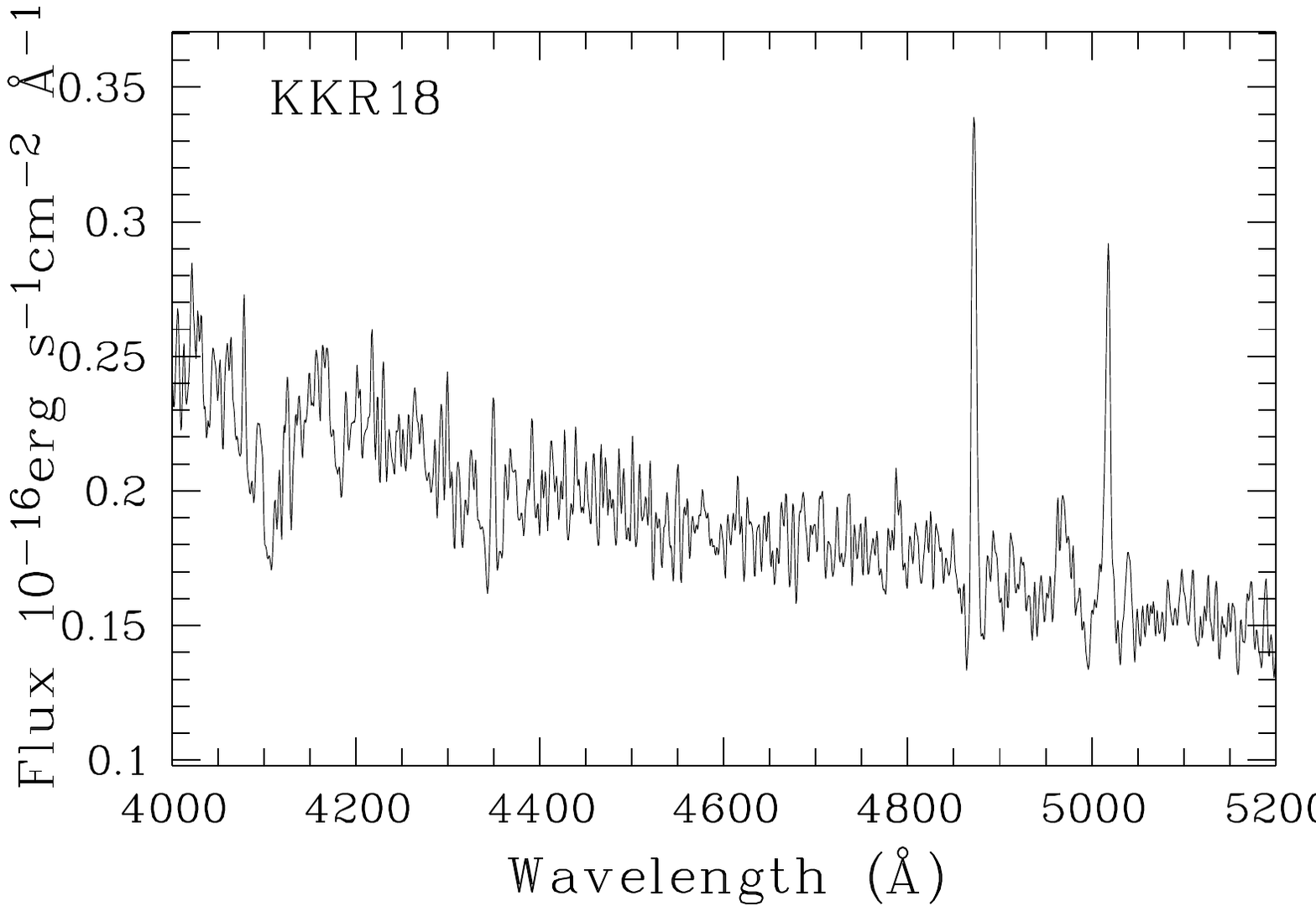}
\caption{BTA-6m spectrum of the dominant emission region in KKR\,18. {\bf Top panel:}  Total spectrum combined of blue and red parts. 
The prominent emission lines arise from [OII]~$\lambda3727$, H$\beta$~$\lambda4861$, [OIII]~$\lambda5007$ and H$\alpha$~$\lambda6563$.
{\bf Bottom panel:} Close-up of blue part of the spectrum. Prominent Balmer absorption features with EWs of $\sim$5~\AA\ 
are traced in the 
2D spectrum along the slit crossing the main body: evidence of the dominance in the light of the underlying relatively young stellar population.}
\label{fig:spectrum}
\end{figure}

\section{Spectra analysis and Oxygen abundance}
\label{sec:oxygen}

The emission lines in the spectrum shown in Fig.~\ref{fig:spectrum} provide a metallicity estimate of 0.04 Solar as will be discussed in this section.
This low value qualifies the galaxy as extremely metal poor (XMP) \citep{2013AJ....146....3S, 2018MNRAS.473.1956I, 2020MNRAS.493..830P, 2021MNRAS.507..944P, 2022MNRAS.510..373A}.

Initial steps, including the establishment of the continuum and the measurement of line fluxes are detailed in \citet{2016MNRAS.463..670P}. Then, to obtain undistorted emission line intensities for the further abundance analysis, we used the iteration procedure from \citet{1994ApJ...435..647I}. It allows us to determine simultaneously the extinction coefficient C(H$\beta$) and the equivalent width of Balmer absorptions EW(abs) in the underlying stellar continuum. 
We note that we assume a two-zone model of an HII region, described by \citet{1990A&AS...83..501S}, in which lines of O$^{+}$ and O$^{++}$ are formed in two zones with different temperatures.

Since the faint auroral line [O{\sc iii}]4363, used for temperature determination in the `direct' method \citep[e.g.][with the most updated formulae]{2006A&A...448..955I} is not detected in our spectrum, we use three empirical calibrators based on the relative fluxes of strong emission lines of Oxygen \citep{2007A&A...462..535Y, 2016MNRAS.457.3678P, 2019A&A...623A..40I}. In the second method, they also use the line fluxes of [N{\sc ii}]6584 or those of the doublet [S{\sc ii}]6716,6731. In addition, we use the  modified `semi-empirical' method \citep{2021MNRAS.507..944P}, in which the temperature T$_{\rm e}$ is derived empirically as suggested in \citet{2007ApJ...665.1115I} and the further parameter estimates explore the formulae from the direct method. 

For the low-metallicity regime of (12+log(O/H) $\lesssim$7.5~dex), one of the most reliable methods is the `strong-line' method by \citet{2019A&A...623A..40I}.
It uses the fit of relations between 12+log(O/H) and parameters R$_{23}$ and O$_{32}$, within the adopted ranges of R$_{3}$ $\geq$ 0.8 and O$_{32}$= $0.65 - 50$. Here, as commonly adopted, 
R$_{2}$ = I([O{\sc ii}]3727)/I(H$\beta$), 
R$_{3}$~=~(I([O{\sc iii}]4959+I[O{\sc iii}]5007)/I(H$\beta$), R$_{23}$ = R$_{2}$ + R$_{3}$. O$_{32}$ is an excitation parameter, defined as I([O{\sc iii}]5007)/I([O{\sc ii}]3727).
The internal scatter of the fit is very small, rms $\sim$0.05~dex. The independent check of their formula on the larger calibration sample in \citet{2021MNRAS.507..944P} results in even the smaller scatter of $\sim$0.04~dex. 

One of the caveats of the method is that its application is limited formally by the constrained range of R$_{3}$ and O$_{32}$ in the calibrator sample. In particular, for a number of XMP dwarfs that are found in voids, spectra fall to the low-excitation regime, with values of O$_{32}$ of 
0.2-0.6. The use of the established formula in this O$_{32}$ regime implies extrapolation outside the formal borders and can incur additional systematics. The same relates to the other O/H calibrators, since they use analogs of parameter O$_{32}$ and the sample of calibration HII regions with the same lower limit on O$_{32}$.

In Table~1 we present 
the measured line fluxes in the spectrum  of KKR\,18 relative to that of H$\beta$ in the 2nd column and then corrected for extinction C(H$\beta$) and the equivalent width of the underlying Balmer absorptions, EW(abs), in the 3rd column. The latter parameters along with the flux F(H$\beta$), corrected for EW(abs), and the equivalent width of the emission H$\beta$ line are presented in the middle of the table. We corrected the measured flux of the line [O{\sc iii}]4959, seriously affected by noise, taking instead its theoretical value of 1/3 the flux in the line [O{\sc iii}]5007. In the bottom of the table we present calculated temperatures in zones of O$^{++}$ and 
O$^{+}$ and the abundance of the respective ions, calculated with the $mse$ method \citep{2021MNRAS.507..944P}. The four last rows present O/H, derived with the strong-line method of \citet{2019A&A...623A..40I} (marked as `s'), the modified semi-empirical method (marked as `mse') \citep{2021MNRAS.507..944P} and with two empirical methods from \citet{2007A&A...462..535Y} (Y07) and \citet{2016MNRAS.457.3678P} (PG16).

In a forthcoming paper (Pustilnik et al., in preparation) we discuss the extension of the \citet{2019A&A...623A..40I} strong-line method to the range of 
O$_{32}$ down to $\sim$0.2, made possible thanks to the spectrum of HII region No.~8 with parameter O$_{32}$ = 0.21 in the XMP galaxy DDO\,68, presented by \citet{2019MNRAS.482.3892A}. Since several other HII regions in DDO\,68 all have 12+log(O/H), derived by the direct method,  in the range 6.96-7.2~dex, No.~8 very likely has O/H closely similar. The estimate of 12+log(O/H) via the strong-line method gives 7.33$\pm$0.05~dex for No.~8. Given its position close to the center of the galaxy, this value is compatible with the metallicity gradient noted by  \citet{2019MNRAS.482.3892A}.

Coming back to our spectrum of KKR\,18, we notice that  all line ratios are close to those of  DDO\,68 HII region No.~8, including the parameter O$_{32}\sim0.24$.
From this similarity, 12+log(O/H)(strong lines) = 7.28$\pm$0.07~dex is taken for the KKR\,18 HII region.
According to the `counterpart' methodology of  \citet{2016MNRAS.457.3678P},  the proximity of the line ratios in KKR\,18 and DDO\,68~HII-region No.~8 from  \citet{2019MNRAS.482.3892A} implies their close values of O/H. Three other estimates of 12+log(O/H) listed at the bottom of Table~2 (mse, Y07 and PG16) are consistent within their uncertainties  with the value of 7.28~dex, adding confidence of the very low metallicity of KKR\,18.

\begin{table}
\centering
\caption{Line intensities and Oxygen abundance of KKR18}
\label{t:Intens}
\begin{tabular}{lcc} \hline
\rule{0pt}{10pt}
$\lambda_{0}$(\AA) Ion                  & F($\lambda$)/F(H$\beta$)&I($\lambda$)/I(H$\beta$)  \\ \hline
3727\ [O\ {\sc ii}]\                           &  3.729$\pm$ 0.357 &   2.446$\pm$ 0.381           \\
4340\ H$\gamma$\                               &  0.073$\pm$ 0.021 &   0.467$\pm$ 0.331           \\
4861\ H$\beta$\                                &  1.000$\pm$ 0.088 &   1.000$\pm$ 0.182           \\
4959\ [O\ {\sc iii}]\                          &  0.302$\pm$ 0.065 &   0.197$\pm$ 0.065           \\
5007\ [O\ {\sc iii}]\                          &  0.917$\pm$ 0.077 &   0.599$\pm$ 0.077           \\
6563\ H$\alpha$\                               &  3.865$\pm$ 0.326 &   2.724$\pm$ 0.388           \\
6584\ [N\ {\sc ii}]\                           &  0.125$\pm$ 0.156 &   0.081$\pm$ 0.156           \\
6716\ [S\ {\sc ii}]\                           &  0.438$\pm$ 0.190 &   0.285$\pm$ 0.190           \\
6731\ [S\ {\sc ii}]\                           &  0.250$\pm$ 0.167 &   0.163$\pm$ 0.167           \\
\hline
C(H$\beta$)\ dex                               & \MC{2}{c}{0.01$\pm$0.11}          \\
EW(abs)\ \AA\                                  & \MC{2}{c}{2.90$\pm$0.64}     \\
F(H$\beta$)$^a$\                               & \MC{2}{c}{1.29$\pm$0.13}      \\
EW(H$\beta$)\ \AA\                             & \MC{2}{c}{ 5.5$\pm$0.3}       \\
\hline
$T_{\rm e}$(OIII)(K)\                          & \MC{2}{c}{21133$\pm$2310}      \\
$T_{\rm e}$(OII)(K)\                           & \MC{2}{c}{15997$\pm$583 }      \\
$N_{\rm e}$(SII)(cm$^{-3}$)\                   & \MC{2}{c}{ 10$\pm$10 ~~} \\
O$^{+}$/H$^{+}$($\times$10$^5$)\               & \MC{2}{c}{1.826$\pm$0.345}     \\
O$^{++}$/H$^{+}$($\times$10$^5$)\              & \MC{2}{c}{0.305$\pm$0.077}     \\
O/H($\times$10$^5$)\                           & \MC{2}{c}{2.132$\pm$0.354}     \\
12+log(O/H)(s)\                                & \MC{2}{c}{7.28$\pm$0.07}     \\
12+log(O/H)(mse)\                              & \MC{2}{c}{7.36$\pm$0.11}        \\
12+log(O/H)(Y07)                               & \MC{2}{c}{7.24$\pm$0.12}     \\  
12+log(O/H)(PG16)                              & \MC{2}{c}{7.24$\pm$0.10}     \\  
\hline
\MC{3}{l}{$^a$ in units of 10$^{-16}$ ergs\ s$^{-1}$cm$^{-2}$}\\
\end{tabular}

\end{table}

\subsection{A Stellar Cluster}
\label{sec:cluster}

\begin{figure}[]
\centering
\includegraphics[width=8.5cm,angle=0,clip=]{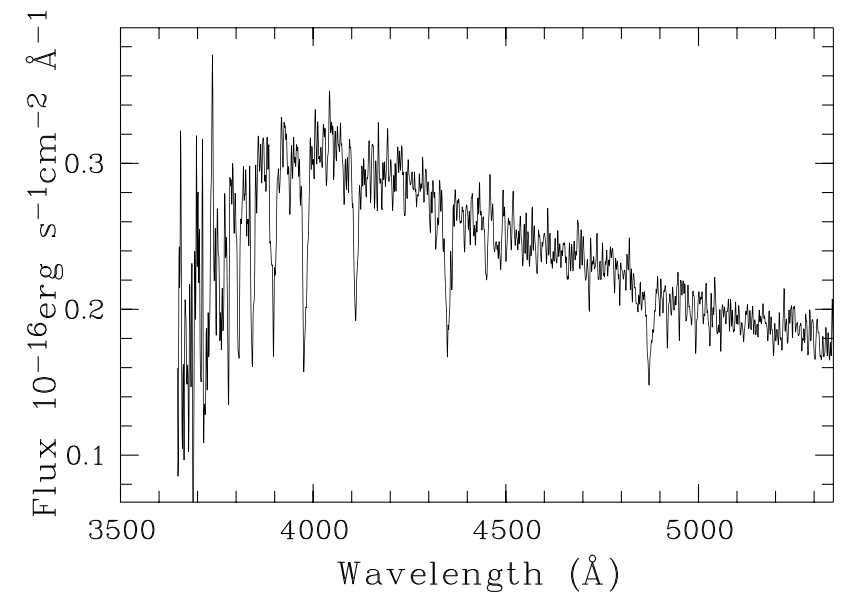}
\caption{
BTA-6m spectrum of the individual young star cluster demonstrating the prominence of the Hydrogen Balmer series in absorption.}
\label{fig:cluster}
\end{figure}

By happenstance, a bright semi-stellar object lies within the slit of the BTA-6m spectral observation.  Its location is indicated by the red arrow in Fig.~\ref{fig:image}.
The extraction of the blue component for this object is shown in Fig.~\ref{fig:cluster}. 
The spectrum is dominated by the strong absorption features of the Hydrogen Balmer series, attributable to A supergiant stars. 

The object was not extracted as a star with the DOLPHOT photometry because it is semi-resolved.  Based on aperture photometry, it has the Vega magnitudes F606W=21.88~mag and F814W=21.58~mag, each with uncertainties $\sim\pm0.1$ mag.  Its location on the KKR\,18 CMD is shown by a star in Fig.~\ref{fig:fullcmd}.
This object is evidently a star cluster with its luminosity dominated by A stars. 
At the assigned distance of KKR\,18, the absolute magnitude of the star cluster is $M_{F814W}=-8.4\pm0.2$ mag, or M$_{\rm V} \sim -8.1$~mag.
We compare the luminosity and color of the cluster with the temporal evolution of these parameters provided in models of Starburst99 \citep{1999ApJS..123....3L}.
For an instantaneous starburst with the Salpeter IMF, $V-I=0.30\pm0.1$ mag and metallicity 5\% Solar, they predict an age for the cluster of $\sim60$~Myr with the $\pm$1$\sigma$ interval due to the color uncertainties of 25 to 220 Myr.

The temporal  evolution of the Balmer line EWs for instantaneous starbursts was presented by \citet{1999ApJS..125..489G}. The observed EWs of the stellar cluster allow one to narrow the age range permitted within the current color uncertainties. 
The measured EW(H$\delta$) = 6.8$\pm$0.5~\AA,
corrected for contamination of continuum ($\sim$25\%) from an adjacent F-supergiant in the slit, 
implies that the real EW(H$\delta$) of the cluster is $\sim$8.5~\AA.
Therefore, an estimate of its age is $\sim$60$\pm$10 Myr, and the total mass of $\sim$10$^{4}~\Msun$.

\section{HI and stellar mass}
\label{sec:mass}

The global 21-cm neutral hydrogen profile of KKR\,18 (Fig.~\ref{fig:HI}) and associated parameters can be found in the All Digital HI catalog in the Extragalactic Distance Database as AGC\,252478 \citep{2015MNRAS.447.1531C, 2018ApJ...861...49H}. 
The HI flux is $F_{HI}=1.55$~Jy.\kms\ and 
the line width at 50\% of mean flux is 50~\kms.
The total HI mass given $d=9.5$~Mpc is $M_{HI}=3.3\times10^7~\Msun$.  Accounting for the accompanying helium with a factor 1.33 the gas mass is $M_{gas}=4.4\times10^7~\Msun$.

The mass in stars can be estimated from the global mid-infrared flux.  \citet{2026arXiv260605780T} report photometry of nearby galaxies at $3.4\mu$m (W1) and $4.6\mu$m (W2) with WISE, the Wide-field Infrared Survey Explorer.  KKR\,18 was not included in that sample, but has subsequently been given attention.  Flux growth curves as a function of major-axis radius for this galaxy are shown in Fig.~\ref{fig:W1}.
The asymptotic apparent magnitudes (total magnitudes to the sky) are W1=16.33$\pm0.04$ mag and W2=17.18$\pm0.03$ mag. 
Reddening at W1 is $A_{W1}=0.005$ mag and is negligible at W2.  At the distance of KKR\,18, the reddening corrected absolute W1 magnitude is $M_{W1}=-13.55$ mag.

\begin{figure}[]
\centering
\includegraphics[width=1.00\linewidth]{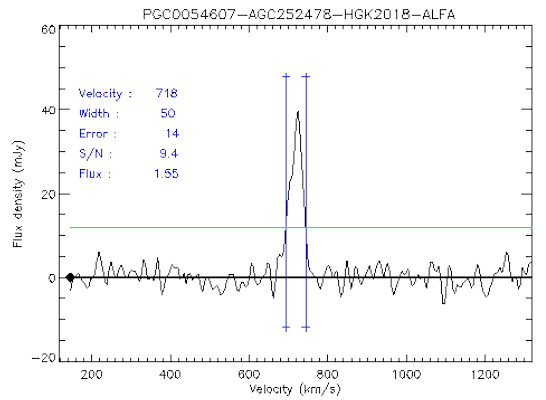}
\caption{
Neutral Hydrogen line profile from Arecibo Telescope observation.}
\label{fig:HI}
\end{figure}

\begin{figure}[]
\centering
\includegraphics[width=1.00\linewidth]{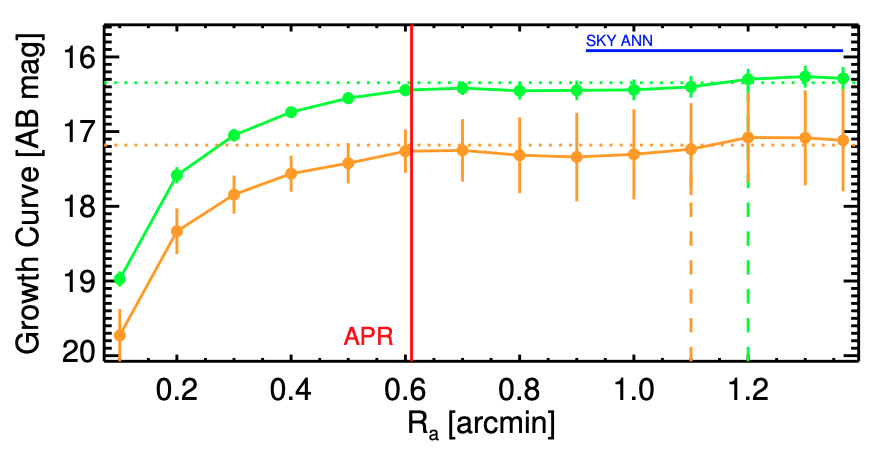}
\caption{WISE flux growth curves with radius. Green: W1 band, orange: W2 band. Asymptotic magnitude values reached at horizontal dotted lines.
}
\label{fig:W1}
\end{figure}

Stellar mass can be approximated from the W1 magnitude with the mass-to-light ratio $\Upsilon_{\star}^{W1}=0.5~\Msun/\Lsun$ \citep{2014MNRAS.445..899C}.  Following \citet{2026arXiv260605780T}, the stellar mass is $M_{\star}=3.0\times10^7~\Msun.$
The ratio of mass in gas to the mass in stars is $M_{gas}/M_{\star}=1.4$.  The total mass in gas and stars is $7.4\times10^7~\Msun$.




 

\section{Discussion}
\label{sec:discussion}

\subsection{Four ``Young'' Candidates}
\label{sec:4cand}

Currently, three other galaxies have been observed with resolved stellar populations at levels that should reveal RGB stars but do so only ambiguously.  These are 
I\,Zw\,18 = PGC\,3167220 \citep{1972ApJ...173...25S},
Leoncino = PGC\,5807231 \citep{2020ApJ...891..181M}, and
Peekaboo = PGC\,5060432 \citep{2023MNRAS.518.5893K, 2025A&A...698L..20K}. 
From their emission spectra, all three, like KKR\,18, are in the XMP class with 12+log(O/H) = (6.99 -- 7.17) dex (or 2--3\% of the Solar value). 
However, unlike these three cases and KKR\,18, galaxies in the XMP class can have well-developed RGB; e.g. Leo\,P \citep{2015ApJ...812..158M, 2015ApJ...815L..17M}.

I\,Zw\,18, the first galaxy to be identified with an RGB deficiency, has been extensively studied with HST \citep{2004ApJ...616..768I, 2007ApJ...667L.151A}, the latter finding a distance from the tentative TRGB of $d=18.2$~Mpc confirmed by a Cepheid distance of $d=19.0$~Mpc \citep{2010ApJ...713..615M}.  
I\,Zw\,18 has been the subject of JWST imaging \citep{2024A&A...689A.146B}, with the brightest 1.5~mag of a weak RGB unambiguously detected.
This galaxy, the most distant example, is also the most luminous ($M_B\simeq -15.3$) and contains the most neutral HI ($1.5$$\times$$10^8~\Msun$) \citep{2012A&A...537A..72L}. The emission line abundance is 12+log(O/H)=7.17, 3\% of Solar \citep{1993ApJ...411..655S}.

Leoncino and Peekaboo are quite similar to each other in global properties.  Leoncino is given an uncertain distance of 12.1~Mpc based on a poorly populated TRGB measurement found with 12 HST orbits \citep{2020ApJ...891..181M}.  
This gives $M_B=-10.6$ mag and with $M_{HI}=1.83\times10^7~\Msun$ the HI totally dominates stars with $M_{HI}/M_{\star}\simeq25$.
The abundance 12+log(O/H)=7.06 is 2.3\% of Solar \citep{2022MNRAS.510..373A}. 

Peekaboo is the nearest at $d=6.8$~Mpc \citep{2023MNRAS.518.5893K}, with attention brought to it by it's proximate velocity of 716~\kms\ \citep{2018MNRAS.478.1611K} but has received the least attention, partly because of confusion from an adjacent 10th magnitude star.  The HI mass is $M_{HI}=1.20\times10^7~\Msun$, the blue magnitude is $M_B\simeq-11.2$ mag and the HI gas to stellar mass ratio is $M_{HI}/M_{\star}\sim 10$.
The abundance 12+log(O/H)=6.99 is 2\% of Solar \citep{2025A&A...698L..20K}. 

KKR\,18 is the second most substantial of the known RGB-deficient galaxies with $M_B=-13.8$ mag. With $d=9.5$~Mpc, KKR\,18 enters the volume-limited sample of Tully et al. 2026 (submitted) of galaxies with TRGB distances $1<d<10$~Mpc and $M_B<-13$ mag.  KKR\,18 is the {\it only} RGB-deficient galaxy in the complete sample of 240 galaxies.  With the statistics of one case, this property occurs 0.4\% of the time.

KKR\,18 is the most unambiguously isolated of the RGB-deficient galaxies, alone in the Local Void 6~Mpc above the Local Sheet and 6~Mpc from NGC\,5055.   Peekaboo galaxy is 370~kpc projected from NGC\,3621 \citep{2023MNRAS.518.5893K} and in a filamentary straggle of dwarfs about 3~Mpc below the Local Sheet in the proximity of the Cen\,A group.  The sharp eye can find these galaxies in the interactive display associated with Fig.~7 in Tully et al. 2026 (submitted).  CMD exist for all the relevant galaxies along this NGC\,3621 filament \citep{2021AJ....162...80A} and the RGB are robust in all except KKR\,18. In the case of Leoncino, the galaxy UGC\,5186 lies at $13^{'}$ in projection and from the HI morphology from VLA mapping is probably interacting \citep{2020ApJ...891..181M}.  
I\,Zw\,18 is clearly interacting with the smaller I\,Zw\,16C. 
The luminous galaxy NGC\,3079 is $\sim2$~Mpc away. 
Of interest, I\,Zw\,18 and Leoncino lie in the same void N.12 in the Nearby Void Galaxy (NVG) catalog  along with two other known XMP galaxies J0926+3343 \citep{2010MNRAS.401..333P} and DDO\,68 \citep{2005A&A...443...91P}.

\subsection{Late Arrival Galaxies}
\label{sec:LAG}

It is familiar in astronomy to refer to a galaxy as ``young" if it has substantial ongoing star formation. 
In the extreme, the acronym VYG (Very Young Galaxy) has been coined to describe galaxies where at least 50\% of stars have been formed in the last Gyr \citep{2018MNRAS.477.1427T, 2020MNRAS.492.1791M}. 
(\citet{2018ApJ...869..152D} have used the term ``late bloomers" to describe apparent late development of massive galaxies at $z\sim0.6$).
However, while the four galaxies described in the previous subsection can qualify as VYG, not all galaxies in the VYG class have ambiguous RGB.  Indeed, \citet{2018MNRAS.477.1427T} discuss VYG within the framework of the standard theory of the growth of galaxies through merging processes.
Recent events might have induced star formation but precursor components might have formed in archaic times.

At the risk of overburdening the lexicography of astronomy, we introduce the term LAG for ``late arrival galaxies", {\it galaxies with depleted RGB}.
The known cases are dwarf galaxies with recent star formation in substantial gas reservoirs and analysis of their emission line spectra place them in the XMP class. 
It can be entertained that, contrary to VYG, there are galaxies with depleted RGB that avoid attention because they are not currently forming many stars.

\citet{2021ApJ...921L...9B} have discussed the limited probability that star formation can begin in dwarf galaxies well after the epoch of cosmic reionization within the $\Lambda$CDM framework.
Their study involved following high resolution hydrodynamical cosmological simulations with attention to present-day dark matter halos with virial masses between $10^9~\Msun$ and $10^{11}~\Msun.$
After reionization, gas cannot collect in dark matter halos with masses below a critical value.  But there is growth in the mass of dark matter halos over cosmic time.  With stochastic variations in growth histories, star formation can be initiated in some halos at late times, even today.  

The simulations by \citet{2021ApJ...921L...9B} offer expectations for the frequency of galaxies with late-time onset of star formation that depends on redshift and baryonic mass.  The inference from the observed properties of the KKR\,18 RGB suggest the serious onset of star formation began within $z<0.5$; i.e., the second half of the age of the Universe.  The estimate for gas plus stellar mass for KKR\,18 is $8\times10^7~\Msun$.  
From the simulations by \citet{2021ApJ...921L...9B}, the likelihood of a galaxy with this baryonic mass commencing star formation at such a late time is very low.  It is a regime that was not fully explored by those authors.  Further theoretical attention must be given to understand if KKR\,18 or other LAGs accord with the $\Lambda$CDM paradigm. 

\section{Conclusions}
\label{conclusions}

Observations of KKR\,18 during an HST SNAP imaging program resolved stars to $\sim1$~mag fainter than the level of the TRGB.  There is a wealth of young stars with ages $20-100$~Gyr and good evidence for AGB stars with ages $0.5 - 1.5$~Gyr. There are RGB stars but far fewer than expected for a galaxy with $M_B\sim-14$.  The RGB stars that are seen lie too close to the limit for reliable photometry in both magnitude and color to give useful age information.  However at ages greater than $\sim3$~Gyr enough low mass stars should have risen up the RGB to contradict the observations.

This paucity of RGB stars in KKR\,18, and in a small number of other galaxies (I\,Zw\,18, Leoncino, Peekaboo), are interpreted as evidence for delayed onset of star formation.  We posit the existence of LAGs, late arrival galaxies.  The distinction between a LAG and VYG, if any, remains to be clarified.  Operationally, a LAG has a deficiency of RGB stars and a VYG is overwhelmingly dominated by young stars.

It is noteworthy that KKR\,18 is extremely isolated, lying in the Local Void.  Morphologically, the galaxy is a dwarf irregular. The current star formation is attenuated from levels $20-100$~Myr ago and the Fig.~\ref{fig:image} image of the galaxy is unremarkable. 

The gold standard for an evaluation of the age of the oldest stars would involve stellar photometry at the level of the main sequence turnoff which probably must await observations with the Extremely Large Telescope. Integrations over a day with James Webb Space Telescope (JWST) could reach a magnitude below the horizontal branch and give the possible detection of RR\,Lyrae stars.  But even a modest observation with JWST could reach 3~mag below the TRGB and settle whether KKR\,18 has a deficient RGB and is truly a LAG.

\bigskip\noindent
Acknowledgement.
Support for this study has been provided by the Space Telescope Science Institute with award HST GO-18070. Additional support for the WISE photometry has come from NASA grant No. 88NSSC18K0424.
Observations with the SAO 6-m telescope BTA were conducted as a part of 
the government contract of SAO RAS approved by Ministry of Science and Higher Education of the Russian Federation. Observations with BTA are also supported by this Ministry (including agreement No.05.619.21.0016, project ID RFMEFI61919X0016).

\bibliography{paper}
\bibliographystyle{aasjournal}

\end{document}